\documentstyle[aps,multicol,psfig]{revtex}

\title{Two-body correlations in Bose condensates}

\author{O. S\o rensen, D.V.~Fedorov, A.S.~Jensen and E. Nielsen}
\address{Institute of Physics and Astronomy,
Aarhus University, DK-8000 Aarhus C, Denmark}

\date{\today}

\begin{document}
\draft

\maketitle

\begin{abstract} 
We formulate a method to study two-body correlations in a condensate
of $N$ identical bosons.  We use the adiabatic hyperspheric approach
and assume a Faddeev like decomposition of the wave function.  We
derive for a f\mbox{}ixed hyperradius an integro-differential equation for
the angular eigenvalue and wave function. We discuss properties of the
solutions and illustrate with numerical results. The interaction
energy is for $N \approx 20$ f\mbox{}ive times smaller than that of the
Gross-Pitaevskii equation.
\end{abstract}


\widetext
\tighten
\begin{multicols}{2}

\paragraph*{Introduction.}

Few-body correlations often express the distinguishing characteristic
features of an $N$-body system \cite{car98}. Two-body correlations are
not only the simplest but in most cases also the most important.
Higher order correlations have a tendency to either strongly confine
spatially or correlate into clusters of particles effectively reducing
the correlations to lower order. Exceptions are the three-body
correlations decisive for the stability of Borromean systems known for
dripline nuclei \cite{rii94}.

Non-correlated mean-field computations of nuclei with the free
two-body nucleon-nucleon interaction produce disastrously wrong
results. Two-body correlations compensating for the short range hard
core repulsion are absolutely necessary \cite{ama98}. For atoms the
effective interaction is also strongly repulsive at shorter distances.
Furthermore for an atomic Bose condensate a short-range two-body
attraction produces diatomic recombination and thereby the
atoms decay out of the condensate \cite{nie99}.  For both nuclei and
molecules correlations are decisive. For nuclei various methods have
been designed to deal with this problem, i.e. Jastrow theory, Bruckner
theory, effective interactions in model spaces \cite{sie87}.

For Bose condensates the Gross-Pitaevskii (non-correlated) mean-field
equation has been the starting point since the first observation of
the condensate in 1995 \cite{dal99}. The wave function does not
include any correlations and the assumed repulsive
$\delta$-interaction has the immediate consequence that the short
range behavior cannot be described even qualitatively correct.
Repulsive zero range potentials are not physically meaningful
\cite{phi97} although sometimes useful in selected Hilbert spaces.
Attractive $\delta$-interactions in three dimensions lead to
divergencies demanding renormalization \cite{fed01,ols01} or a change
of boundary conditions \cite{dem88}.  When two-body bound states
appear even dimer condensates may be possible \cite{pri00}.

To include correlations we must necessarily go beyond the mean-field
Hartree-Fock-Boguliubov approximation.  Then finite range potentials
with realistic features can as well be used as the starting point in
the theoretical formulation. The other crucial ingredient is the
degrees of freedom or, equivalently, the Hilbert space.  An
interesting formulation was recently introduced in terms of
generalized hyperspherical coordinates and an adiabatic expansion with
the hyperradius as the adiabatic coordinate \cite{boh98}. Still only a
zero-range interaction was used with the lowest hyperspherical angular
wave function. These are very crude approximations, since the
expansion in the hyperspherical basis necessarily must be extremely
slowly converging for large(r) hyperradii.

The purpose of this letter is to go a step further and establish the
optimum equation to determine two-body correlations. We shall derive a
practical and realistic equation applicable to a system of $N$
identical bosons.  We shall allow general two-body interactions and in
particular both attractive and repulsive finite range potentials.  The
resulting equation may be considered as an alternative to the
Gross-Pitaevskii equation, but allowing two-body correlations.

\paragraph*{Theory.}

The system of $N$ identical interacting bosons of mass $m$ may be
described by the coordinates $\vec{r}_{j}$ or CM ($\vec R$) and the
$k=1,2,\ldots,N-1$ Jacobi coordinates \cite{bar99}
\begin{eqnarray}  \label{e2}
\vec{\eta}_k&=&\sqrt{{N-k\over N-k+1}}
\biggl(\vec{r}_{N-k+1}-{1\over N-k}\sum_{j=1}^{N-k}\vec{r}_j\biggr) \; .
\end{eqnarray}
The hyperspherical coordinates are given by the hyperradius $\rho$ and
the hyperangles $\alpha_k\in[0,\pi/2]$
\begin{eqnarray} \label{e4}
\rho_l^2=\sum_{k=1}^l\eta_k^2 , \; \rho^2\equiv 
\sum_{i<j=1}^N \frac{r_{ij}^2}{N} , \;
\eta_k=\rho_k \sin\alpha_k  ,
\end{eqnarray}
where $r_{ij}=|\vec{r}_{i}-\vec{r}_{j}| \equiv \sqrt{2} \rho \sin
\alpha_{ij}$, and we distinguish between one and two indices on
$\alpha$.  The remaining $2(N-1)$ angles are the directions of the
$N-1$ $\vec\eta_k$-vectors.  All the angles together, including
the $\alpha$'s, are denoted $\Omega$.

Removing the center of mass motion the intrinsic hamiltonian $\hat H$
for a trap of angular frequency $\omega$ becomes
\begin{eqnarray}
\hat H=\hat T+\sum_{i<j=1}^N({1\over2}m\omega^2{1\over N}r_{ij}^2+V_{ij})
 \; ,
\end{eqnarray}
where $V_{ij}=V(r_{ij})$ is the two-body interaction and the internal
kinetic energy $\hat T$ is of the form \cite{bar99}
\begin{eqnarray} \label{e14}
\hat T&=&-{\hbar^2\over 2m}\biggl[{1\over\rho^{3N-4}}
{\partial\over\partial\rho}
\rho^{3N-4} {\partial\over\partial\rho}-{1\over\rho^2}
\hat\Lambda^2\biggr] \; , \\ \label{e18}
\hat\Lambda^2 &=& - {\partial^2\over\partial\alpha^2}+
{3N-9-(3N-5)\cos 2\alpha\over\sin 2 \alpha}{\partial\over\partial\alpha}
+ D \; ,
\end{eqnarray}
where $D$ only contains derivatives with respect to angles different
from $\alpha \equiv \alpha_{12}$.  The hamiltonian is then
\begin{eqnarray}
\hat H&=&\hat H_\rho+{\hbar^2\over2m\rho^2}\hat h_\Omega \;\; , \;\; 
\hat h_\Omega=\hat\Lambda^2+\sum_{i<j=1}^Nv_{ij}\; , \label{e20} \\
\hat H_\rho&=&{1\over2}m\omega^2\rho^2-{\hbar^2\over2m}{1\over\rho^{3N-4}}
{\partial\over\partial\rho}\rho^{3N-4} {\partial\over\partial\rho} 
\; , \label{e22}
\end{eqnarray}
where $v_{ij}=V_{ij} 2m\rho^2 /\hbar^2$ is a dimensionless potential.

The total wave function $\Psi$, obeying $\hat H\Psi=E\Psi$, is for
each $\rho$ expanded as
\begin{eqnarray} \label{e24}
\Psi&=\rho^{-(3N-4)/2}\sum_nf_n(\rho)\Phi_n(\rho,\Omega) \; ,
\end{eqnarray}
where $f_n(\rho)$ are the expansion coefficients on the solutions
$\Phi_n(\rho,\Omega)$ to the angular eigenvalue equation
\begin{eqnarray} \label{e26}
(\hat h_\Omega-\lambda_n)\Phi_n(\rho,\Omega)=0 \; .
\end{eqnarray}
The coupled set of radial equations are then \cite{bar99}

\begin{eqnarray}
&&\bigg(-{d^2\over d\rho^2} + {\lambda_n(\rho)\over\rho^2} +
{2m(U(\rho)-E)\over\hbar^2}\bigg)f_n = \nonumber\\
&&\sum_{n'}\big(2Q_{nn'}^{(1)}(\rho){d\over d\rho}+Q_{nn'}^{(2)}(\rho)\big)
f_{n'} \; ,
\label{e28} \\ &&
Q_{nn'}^{(i)}(\rho) \equiv \int d\Omega
[\Phi_n(\rho,\Omega)]^*{\partial^i\over\partial\rho^i}\Phi_{n'}(\rho,\Omega)
\label{e32}
\end{eqnarray}
with $U(\rho)$ from the external trap and centrifugal barrier 
\begin{eqnarray} \label{e29}
U(\rho)&=&{1\over2}m\omega^2\rho^2+{\hbar^2\over8m\rho^2}(3N-4)(3N-6) \; .
\end{eqnarray}
Then $\Phi_n$ is decomposed in Faddeev components $\phi_{ij}$ as
\begin{eqnarray} \label{e34}
\Phi_n(\rho,\Omega)&=&\sum_{i<j=1}^N\phi_{ij}^{(n)}(\rho,\Omega) \; .
\end{eqnarray}
We assume that each two-particle amplitude $\phi_{ij}$ can be
described by $s$-waves and therefore only depends on $\rho$ and the
distance $r_{ij}$ \cite{jen97}.  This does not exclude higher partial
waves in the total wave function, e.g.  expressing the ``$12$''
$s$-wave component in relative ``$34$''-coordinates requires non-zero
angular momenta.  The boson symmetry implies that
$\phi_{ij}^{(n)}(\rho,\Omega) = \phi^{(n)}(\rho,\alpha_{ij}) =
\phi(\alpha_{ij})$ are identical functions of the different
coordinates $\alpha_{ij}$.

The angular eigenvalue $\lambda$ in eq.(\ref{e26}) is given by
\begin{eqnarray}  \label{e38}
\lambda={\langle\Phi|\hat h_\Omega|\Phi\rangle\over\langle\Phi|\Phi\rangle}
 = {\langle\phi_{12}|\hat h_\Omega|\sum\phi_{kl}\rangle\over
 \langle\phi_{12}|\sum\phi_{kl}\rangle} \; .
\end{eqnarray}
We now vary $\phi^*_{12}$ for each $\rho$ and find the condition for
extremum of $\lambda$.  From eqs.(\ref{e20}) and (\ref{e38}) we then
obtain the integro-differential equation
\begin{eqnarray}  \label{e40} 
 \int d\Omega\sum_{k<l}\bigg[(\hat\Lambda^2-\lambda)
\phi_{kl}+v_{kl}\sum_{m<n}\phi_{mn}\bigg] = 0 \; ,
\end{eqnarray}

The integrals in eq.(\ref{e40}) involve at most six particle
coordinates. By appropriate choices \cite{smi77} of Jacobi systems
these can be expressed in terms of the five vectors
$\vec\eta_{N-1},\ldots,\vec\eta_{N-5}$. One variable
($\alpha\equiv\alpha_{12}$) is the argument of the variational
function $\phi_{12}$ and not an integration variable. Furthermore 7
variables only enter in the phase space and can be integrated
analytically leaving at most five-dimensional integrals. With
$v(\alpha) \equiv v_{12}({\sqrt 2} \rho \sin\alpha)$ we rewrite
eq.(\ref{e40})
\begin{eqnarray}  \label{e42} 
 \Big(\hat\Lambda^2 + v(\alpha)-\lambda\Big)
\phi(\alpha)+\int d\tau G(\tau,\alpha) = 0 \; , 
\end{eqnarray}
where $\tau$ denotes the remaining 5 variables and $G$ is a definite
homogeneous linear combination of $\phi(\alpha_{ik}(\tau,\alpha))$.

The integrals in eq.(\ref{e42}) reduce to at most two dimensions when
the two-body interaction-range $b$ is much less than the
characteristic length $\rho$ of the system.  To illustrate we use a
gaussian potential $V(r_{ij}) = V_0\exp(-r_{ij}^2/b^2)$, but the
results depend mainly on $a_p \equiv \sqrt\pi mb^3V_0/4\hbar^2$, i.e.
the Born-approximation of the scattering length $a_s$.  When
$mb^2V_0 \ll \hbar^2$ we have $a_p \approx a_s$.  For $b \ll
\rho$ we get, almost independent of the potential,
\begin{eqnarray} 
&&\int d\tau G(\tau,\alpha)\simeq   \tilde v(\alpha)  
\big(1 + 2 g(\alpha) \big)\phi(\alpha)  \nonumber \\  \label{e48} 
&&\qquad + \hat R(\alpha) +  \tilde v(\alpha)
\big(1 + g(\alpha) \big)\phi(\alpha=0)  \; , \\  \label{e53} 
&&\tilde v(\alpha) = \sqrt{2\over\pi}\; \frac{\Gamma ({3(N-2)\over2})}
{\Gamma({3(N-3)\over2})}\;  {a_p\over\rho} \;\frac{(N-2)(N-3)}{\cos^3\alpha} \; , \\
&& g(\alpha) = {32\over3\sqrt3}
\Theta(\tan\alpha<\sqrt3)\frac{\cos^{3N-11}\beta}{N-3} \; .
\end{eqnarray}
Here $\Gamma$ is the gamma function,
$\sin\beta=\tan\alpha/\sqrt3$, $\Theta$ is the truth function, and
$\hat R$, a functional of $v$ and $\phi$ and a function of $\alpha$, arises
from rotations of interaction and wave function from one set of
Jacobi coordinates to another.

The $\delta$-interaction with a constant wave function, labeled $K=0$,
leads to the eigenvalue \cite{boh98} 
\begin{eqnarray} \label{e58}
 \lambda_{K=0} = 
  \sqrt{2\over\pi} \; \frac{{\Gamma({3(N-1)\over2})}}
{{\Gamma({3(N-2)\over2})}}\; N (N-1) \; {a_p\over\rho} \; ,
\end{eqnarray}
which for large $N$ coincides with the term $\tilde v(\alpha=0)$ in
eq.(\ref{e48}). The interaction dependent part of the strength is
collected in the parameter $a_p$.

The limit of $\lambda_{K=0}$ is not, even for large $N$, obtained by
solving eqs.(\ref{e42}) and (\ref{e48}), since the highest power of
$N$ is found in one of the rotated terms in $\hat R$ proportional to
$N^2 \tilde v(\alpha)$.  In the zero-range approximation the two
non-local terms in eq.(\ref{e48}) proportional to $\phi(0)$ are very
sensitive to the initial two-body potential $v(\alpha)$ which is very
large and of very short range already for moderate values of
$\rho$. These terms are therefore crucial for the developement of
correlations.

\paragraph*{Quantitative behavior.}

We now solve numerically eq.(\ref{e42}) using eq.(\ref{e48}) with the
realistic parameter set in \cite{boh98} for the condensate of
$^{87}Rb$-atoms. For numerical illustration we choose $N=10,20,30$.

\begin{figure}
\begin{center}
\psfig{figure=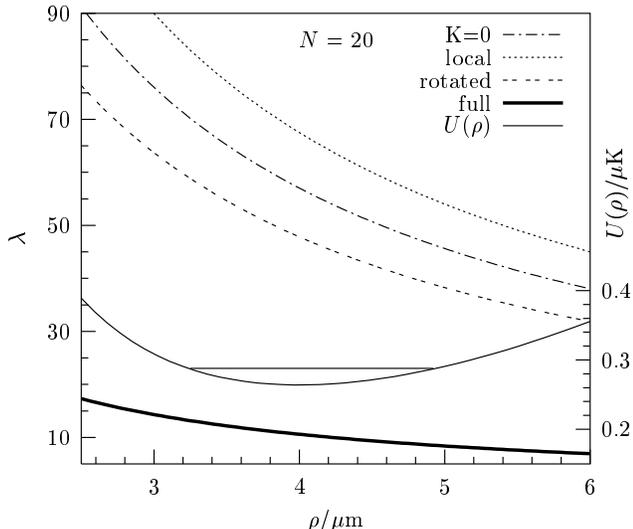,%
bbllx=5.8cm,bblly=19.4cm,bburx=15.8cm,bbury=27.6cm,width=8.6cm}
\end{center}
\vspace*{-0.7cm}
\caption{The lowest angular eigenvalue $\lambda$ obtained from
eq.(\ref{e42}) as function of $\rho$ for $20$ $^{87}Rb$-atoms. The
parameters are $a_p=5.29$ nm, $b=a_p/10$, $\nu = \omega /(2\pi)=
200$~Hz, ($b_t=763$~nm). The decreasing curves show the energy in
eq.(\ref{e58}), the energy including only the local terms of
eq.(\ref{e48}), local plus $\hat R$-terms (rotated) and the full
energy with all terms. Shown is also $U(\rho)$ (parabolic curve) in
eq.(\ref{e29}) along with the free center of mass energy
$3(N-1)\hbar\omega /2$ (horizontal line). }
\label{fig1}
\end{figure}

The lowest angular eigenvalues for $N=20$ are shown in Fig.~\ref{fig1}
in the relevant range of hyperradii selected by the external trap. The
minimum of $U(\rho)$ is located near $\rho\simeq b_{t} \sqrt{3N /2}$
where $b_t=\sqrt{\hbar/(m\omega)}$ is the trap length.  Inclusion of
the interaction terms, expressed in eq.(\ref{e28}) by $\lambda$
through the effective radial potential
$U(\rho)+\hbar^2\lambda/2m\rho^2$, push the minimum outwards. For the
repulsive interaction the angular eigenvalues decrease with $\rho$ due
to an increasing average two-body distance.

We first only include the local terms ($\propto \phi(\alpha)$) in
eq.(\ref{e48}). The angular energy in Fig.~\ref{fig1} is then larger
than the $\lambda_{K=0}$ value, because these terms are
repulsive. Including also the $\hat R$-terms in eq.(\ref{e48}) leads
to rather similar results, where however, the correlations are
exploited giving a lower energy.  Finally, inclusion of all terms
reduces the angular energy to only $19\%$ of the $\lambda_{K=0}$
value.

\begin{figure}
\begin{center}
\psfig{figure=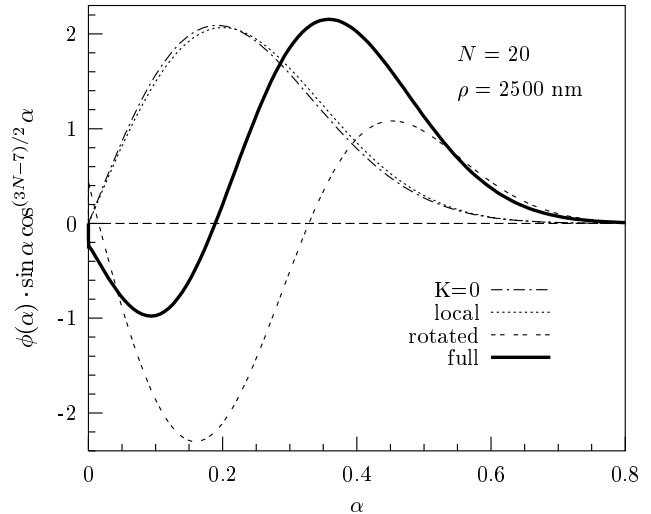,%
bbllx=5.8cm,bblly=19.4cm,bburx=15.8cm,bbury=27.6cm,width=8.5cm}
\end{center}
\vspace{-0.8cm}
\caption{The wave function $\phi$ multiplied by the square root of the
  volume element, $\sin\alpha\cos^{(3N-7)/2}\alpha$, as function of
  $\alpha$ for $\rho = 2500 $ nm ($= \sqrt{20}\cdot 559 $ nm) for the
  $\lambda$'s in Fig.~\ref{fig1}.}
\label{fig2}
\end{figure}

The corresponding angular wave functions in Fig.~\ref{fig2} are all
zero at $\alpha=0$ due to the vanishing volume element.  The almost
discontinous behavior at this point of the full and rotated angular
wave function is caused by the very short range of the initial
repulsive two-body potential $v(\alpha)$. The $K=0$ wave function has
no nodes and the local terms maintain this structure but marginally
shifted to larger $\alpha$-values by the repulsive potential. The
rotated terms introduce one node indicating the substantial
restructuring due to correlations. The higher kinetic energy is more
than compensated by the correlations build up to avoid the repulsion
at short distance.  The full solution maintains qualitatively this
behavior, but now the probability is shifted to larger $\alpha$ or
equivalently to larger distances between each pair of particles.

The terms proportional to $\phi(0)$ decreases the angular energy
drastically.  This is consistent with the node in the wave function
which lowers the angular energy because then large $|\phi(0)|$ results
in smaller $\lambda$.  More oscillations allow constructive
interference which in turn substantially lowers the potential energy.
The result is the increasing deviation of $\lambda$ from
$\lambda_{K=0}$ for decreasing $\rho$, see Fig.~\ref{fig1}.

\begin{figure}
\begin{center}
\psfig{figure=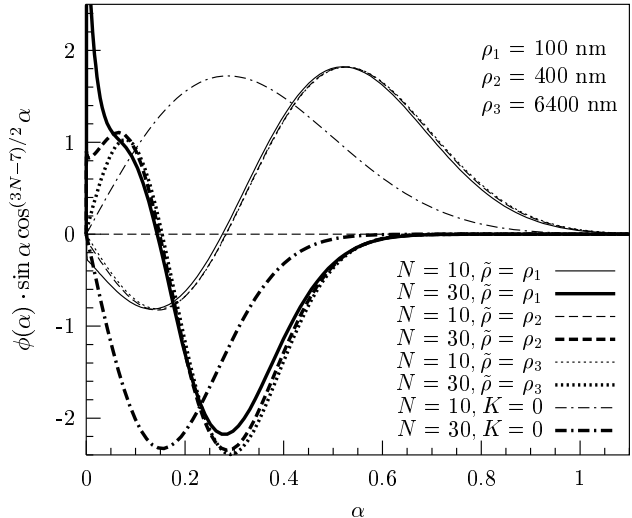,%
bbllx=5.8cm,bblly=19.4cm,bburx=15.8cm,bbury=27.6cm,width=8.6cm}
\end{center}
\vspace{-0.8cm}
\caption{The same as Fig.~\ref{fig2} with full solutions for
  hyperradii $\tilde \rho = \rho / \sqrt N$ = 100 nm, 400 nm, 6400 nm and
  $N=10, 30$.}
\label{fig3}
\end{figure}

The $N$ and $\rho$ dependences of the angular wave functions are shown
in Fig.~\ref{fig3}. The $\sqrt N$ scaling follow the $N$ dependence of
the potential minimum of $U(\rho)$, see Fig.~\ref{fig1}.  The strong
variation for small $\alpha$ for $N=30$ disappears with increasing
$\rho$.  The same tendency, but less pronounced, is found for $N=10$.
The size of the angular wave functions for $\alpha \approx 0$ decreases
with increasing $\rho$.  In the relevant parameter interval defined by
the trap the $\rho$-dependence of the angular wave function is weak
except at small $\alpha$ corresponding to distances inside the
two-body potential.  The corresponding angular eigenvalue $\lambda$
relative to $\lambda_{K=0}$ is essentially independent of $\rho$,
i.e. 0.33, 0.19 and 0.16 for $N=10,20$ and $30$, respectively. This
confirms numerically our conjecture that the $K=0$ behavior obtained
in \cite{boh98} with a $\delta$-interaction is not approached with
increasing $N$ at large distances.

The total energies obtained by solving the radial equation in
eq.(\ref{e28}) reflect the relative sizes of the angular eigenvalues
in Fig.~\ref{fig1}.  The radial wave functions $f_n(\rho)$ resemble
the $K=0$ solutions. However, it is essential to understand that even
if the effective radial potential only deviates insignificantly from
the $K=0$ potential the corresponding angular wave functions may
differ enormously, i.e. none versus many oscillations. Thus effects of
correlations may easily be strong without showing up in the total
energy and perhaps also not even significantly in the interaction
energy.

The Gross-Pitaevskii mean-field solution does not remove the (small)
spurious contribution from the center of mass motion. Still we compare
directly and find that our total energy only is slightly smaller than
this total mean-field energy. However, our interaction energy,
excluding the external trap energy, is about $3$, $5$, and $7$ times
smaller for $N=10$, $20$, and $30$, respectively.  This reflects the
huge difference in structure between the mean-field and correlated
solutions.

The distance between two particles is $r_{ij} = {\sqrt 2} \rho \sin
\alpha_{ij}$ and therefore, the mean square ``radius'' $\langle \phi
(\alpha) | \sin ^2 \alpha | \phi (\alpha) \rangle $ obtained with
normalized $\phi$, quantifies this distance for each $\rho$. For the
$\rho$-independent $K=0$ wave function we find $1/(N-1)$ analytically
whereas our full wave function with the correlations for $N=20$ gives
the much larger distance of about $2.4 / (N-1)$ roughly independent of
$N$ and $\rho$, see Figs.~\ref{fig1} and \ref{fig2}. The full mean
square radius is now obtained by multipliplying $\langle \sin ^2
\alpha \rangle $ with $2 \langle \rho^2 \rangle \approx 3 (N-1) b^2_t$
found as the minimum of $U$ in Fig.~\ref{fig1} or as the harmonic
oscillator expectation value. Thus by increasing $N$ the decreasing
two-particle average distance for fixed $\rho$ is precisely
compensated by the increasing position of the minimum of the potential.

The result is $\langle r_{ij}^2/b^2_t\rangle = $ 3 for $K=0$ and 6.83,
7.30 and 7.31 for the full wave function for $N=10,20,30$,
respectively. These values can be compared to the Gross-Pitaevskii
results through the harmonic oscillator relation $(N-1) \langle
r^2_{ij} \rangle = 2 N (\langle r^2_i \rangle - \langle R^2 \rangle)$,
where $ N \langle R^2 \rangle = 3 b^2_t/2$ is the center of mass
expectation value for the mean square radius. With the mean-field
result numerically computed we obtain $\langle r^2_{ij} / b^2_t
\rangle \approx 3.1$ almost independent of $N$, i.e. almost equal to
the $K=0$ result but much smaller than for the correlated solution.
Thus the $K=0$ assumption do not even approximately describe two-body
correlations. In the mean-field approximation these correlations are
by definition completely absent and the average distance between pairs
of particles is therefore very small.  The accurate detailed behavior
can only be obtained by explicit inclusion of correlations.

\paragraph*{Perspectives.}

In conclusion, we have variationally derived a linear one-dimensional
integro-differential equation describing $N$ identical bosons in a
trap. The equation involves five-dimensional integrals, a tremendous
simplification from the original $N$-body problem.  We assume the
Faddeev angular decomposition of the wave function and use the
hyperradius as the adiabatic coordinate. A further reduction to at
most two-dimensional integrals is achieved for short range
interactions for the non-local parts of the equation. The interaction
may be attractive and of finite range with bound states and with both
signs of the scattering length. This opens the possibility of
realistic computation of the diatomic recombination rate which in the
present terminology is a process originating from the first excited
adiabatic state (condensate) ending in the lowest adiabatic (diatomic
bound) state.  This process is essential for the stability of atomic
Bose condensates. Our equation is an alternative to the
Gross-Pitaevskii equation, but apparently both features of mean-field
and few-body correlations are now included in a unified
approach. Other interesting perspectives are extensions to three-body
correlations, to one and two space dimensions, to non-identical bosons,
to fermion systems and to systems of mixed symmetry.

\end{multicols}


\begin{thebibliography}{99}


\bibitem{car98} J. Carlson and R. Schiavilla, Rev. Mod. Phys. 70, 743 (1998).

\bibitem{rii94} K.~Riisager, Rev. Mod. Phys. 66, 1105 (1994).

\bibitem{ama98} J.E. Amaro et al., Phys. Rev. C 57, 3473 (1998).

\bibitem{nie99} E. Nielsen and J.  Macek, Phys. Rev. Lett. 83, 1566 (1999).

\bibitem{sie87} P.J. Siemens and A.S. Jensen, Elements of Nuclei, 
Addison-Wesley, 1987.

\bibitem{dal99} F. Dalfovo et al., Rev. Mod. Phys. 71, 463 (1999).

\bibitem{phi97} D.R. Phillips and T.D. Cohen, Phys. Lett. B 390, 7 (1997).

\bibitem{fed01} D. Fedorov and A.S. Jensen, Phys. Rev. A 63, 063608 (2001).

\bibitem{ols01} M. Olshanii and L. Pricoupenko, arXiv: cond-mat/0101275.

\bibitem{dem88} Yu.N. Demkov and V.N. Ostrovskii, Zero-range
potentials and their applications in Atomic Physics (Plenum, New York,
1988).

\bibitem{pri00} L. Pricoupenko, arXiv: cond-mat/0006263.

\bibitem{boh98} J.L. Bohn et al., Phys. Rev. A 58, 584 (1998).

\bibitem{bar99} N. Barnea, Phys. Lett. B 446, 185 (1999); Phys. Rev. A
59, 1135 (1999).

\bibitem{jen97}  A.S. Jensen et al., Few-Body Systems 22, 193 (1997).

\bibitem{smi77} Yu. F. Smirnov et al., Sov. J. Part. Nucl. 8(4), 344 (1977).

\end{thebibliography}
\end{document}